\documentclass[
reprint,
aps,
superscriptaddress,
nofootinbib,
nobibnotes,
onecolumn]{revtex4-2}

\usepackage{xcolor}
\usepackage{graphicx}
\graphicspath{{./fig/}}
\usepackage{amsmath,amssymb,amsfonts,mathrsfs,bm}
\usepackage{slashed}
\usepackage{subfig}
\usepackage{float}
\usepackage[%
colorlinks=true,
linkcolor=blue,
citecolor=blue,
]{hyperref}

\begin{document}

\title{Detecting the dark sector through scalar-induced gravitational waves}

\author{Xiao-Bin Sui}
\email{suixiaobin21@mails.ucas.ac.cn}
\affiliation{School of Fundamental Physics and Mathematical Sciences, Hangzhou Institute for Advanced Study, University of Chinese Academy of Sciences (HIAS-UCAS), Hangzhou 310024, China}
\affiliation{CAS Key Laboratory of Theoretical Physics, Institute of Theoretical Physics, Chinese Academy of Sciences, Beijing 100190, China}
\affiliation{University of Chinese Academy of Sciences, Beijing 100049, China}

\author{Jing Liu}
\email{liujing@ucas.ac.cn}
\affiliation{International Centre for Theoretical Physics Asia-Pacific, University of Chinese Academy of Sciences, Beijing 100190, China}
\affiliation{University of Chinese Academy of Sciences, Beijing 100049, China}

\author{Xing-Yu Yang}
\email{xingyuyang@kias.re.kr}
\affiliation{Quantum Universe Center (QUC), Korea Institute for Advanced Study, Seoul 02455, Republic of Korea}

\author{Rong-Gen Cai}
\email{cairg@itp.ac.cn}
\affiliation{School of Physics Sciences and Technology, Ningbo University, Ningbo, 315211, China}
\affiliation{CAS Key Laboratory of Theoretical Physics, Institute of Theoretical Physics, Chinese Academy of Sciences, Beijing 100190, China}
\affiliation{University of Chinese Academy of Sciences, Beijing 100049, China}
\affiliation{School of Fundamental Physics and Mathematical Sciences, Hangzhou Institute for Advanced Study, University of Chinese Academy of Sciences (HIAS-UCAS), Hangzhou 310024, China}

\begin{abstract}
We investigate the evolution of cosmological scalar perturbations in the case that the background radiation is weakly coupled to a light scalar field $\phi$. 
The light scalar $\phi$ is a homogeneous background field with a large initial value.
In the radiation-dominated Universe, the coupling term introduces an effective mass to $\phi$ and the background ultra-relativistic particles. The oscillations of $\phi$ result in the periodic change of the equation of state parameter and the sound speed, which provides a novel mechanism to amplify subhorizon scalar perturbations through parametric resonance.  
The amplification of scalar perturbations leads to a stochastic gravitational-waves background~(SGWB) expected to be observed by multiband gravitational wave observers. The observation of the SGWB helps to determine the initial value of $\phi$ and the coupling strength of the interaction.
This mechanism is generally applicable to the interactions that introduce an effective mass, and we take the interaction between $\phi$ and electrons as a concrete example to illustrate the result. We find that under the condition that the coupling coefficient $\lambda=10^{-16}$ and the initial value $\phi_i=10^{18}$ GeV, the resulting SGWB spectrum is expected to be observed by the future observers including LISA, $Taiji$, DECIGO and BBO.

\end{abstract}

\maketitle

\section{Introduction}

The detection of gravitational waves (GWs) in 2015 marked the beginning of GW cosmology and astronomy~\cite{LIGOScientific:2016aoc,Cai:2017cbj,Bian:2021ini}.
In addition to GWs produced by individual merger events of compact binaries, the stochastic gravitational-wave backgrounds~(SGWBs) have also received extensive attention recently as an important scientific target of LIGO/VIRGO/KAGRA~\cite{KAGRA:2021kbb}, Taiji~\cite{Ruan:2018tsw}, TianQin~\cite{TianQin:2015yph} and LISA~\cite{Barausse:2020rsu}, which are generated by the superposition of GWs from numerous uncorrelated sources. 
The SGWBs of astrophysical origin are produced from the population of astrophysical sources such as binaries of black holes, neutron stars, and white dwarfs. 
The violent physical processes in the early Universe, such as vacuum quantum fluctuations during inflation, phase transition~\cite{Liu:2021svg,Ellis:2019oqb,Khlopov:1998nm,Baker:2021nyl,Cai:2024nln,Flores:2024lng}, and primordial curvature perturbations~\cite{Baumann:2007zm,Saito:2008jc, Ananda:2006af, Kohri:2018awv,Cai:2018dig,Cai:2019amo,Cai:2019cdl,Domenech:2024rks,Pi:2024jwt,Yu:2024xmz,Yu:2023jrs,Martin:2019nuw, Martin:2020fgl}, are also expected to generate observable SGWBs which carries abundant information of new physics and the early history of the Universe.
The constraints from the upper bound of SGWBs can be found in Refs.~\cite{NANOGrav:2021flc,Xue:2021gyq,Romero:2021kby}.

In particular, there has been increasing attention on the second-order GWs induced by scalar perturbations~\cite{Inomata:2020yqv,Inomata:2018epa,Inomata:2019ivs,Inomata:2019zqy,Yuan:2019udt,Cai:2019elf,Yuan:2019wwo,Inomata:2019yww,Domenech:2019quo,Fu:2019vqc,Bhattacharya:2023ysp,Domenech:2021ztg}.
Many well-motivated inflationary models predict strong curvature perturbations at small scales which also coincide well with the observations of the cosmic microwave background and large-scale structure at large scales~\cite{Yokoyama:1998pt,Garcia-Bellido:2016dkw,Cheng:2016qzb,Cheng:2018yyr,Xu:2019bdp,Mishra:2019pzq,Liu:2020oqe,Fu:2020lob,Ragavendra:2020sop}. 
The generation of large-amplitude curvature perturbations is realized mainly through quantum fluctuations of a non-attractor phase during inflation, including ultra-slow-roll inflation~\cite{Di:2017ndc,Liu:2020oqe,Kristiano:2022maq,Fu:2020lob,Choudhury:2023hfm}, sound speed resonance~\cite{Cai:2018tuh,Cai:2019jah} and small structures of the inflationary potential~\cite{Cai:2019bmk,Chen:2024gqn}. 
In this work, we consider a novel scenario that curvature perturbations are amplified by parametric resonance at subhorizon scales after inflation, which originate from the weak coupling between the background plasma and an additional light scalar field.
Observing such scalar-induced GWs could serve as a novel probe for detecting the dark sector beyond the Standard Model~(SM) of particle physics.

Despite the great success of the SM of particle physics in understanding nature, it still falls short in addressing major cosmological problems, with dark energy and dark matter remaining unaccounted for within the SM.
The cold dark matter model has successfully described the formation of large-scale structures, however, there are still challenges in predicting simulations on sub-galactic scales.
To address this issue, a potential candidate for dark matter has been proposed, referred to as wave dark matter or fuzzy dark matter~\cite{Hui:2021tkt,Annulli:2020lyc}.
The masses of this kind of DM are as low as about $10^{-22}$ eV~\cite{Rogers:2020ltq,Dalal:2022rmp,Amorisco:2018dcn,Bar:2019bqz} and the de Broglie wavelengths are larger than the small structures in the galaxies.
Such a light scalar field could even be considered as a potential candidate for dark energy~\cite{Copeland:2006wr,Bamba:2012cp}.
The theories of the dilaton~\cite{Damour:1994zq} and the dark sector~\cite{Nelson:2011sf} also predict the existence of a large number of light scalar fields.
In addition, these light scalar fields may also exist alongside the inflationary field during inflation, such as the spectator field~\cite{Ozsoy:2014sba,Goolsby-Cole:2017hod,Yu:2023ity,Romano:2008rr,Carney:2012pk}, and curvaton~\cite{Enqvist:2001zp, Pi:2021dft,Lyth:2002my,Bartolo:2003jx,Sasaki:2006kq}, producing interesting effects in the subsequent evolution of the Universe~\cite{Cui:2023fbg}. 
After inflation, these light fields exist in the form of a static homogeneous background field and start to oscillate when their effective masses exceed the Hubble parameter. We mainly focus on this case in this work.

The presence of scalar fields in the early Universe has led to a wealth of intriguing phenomena.
Most theories predict the interactions between these scalar fields and the SM. However, generally speaking, these interactions are very weak and difficult to detect.
In this paper, we investigate cosmological probes of a light scalar field through the amplification of curvature perturbations after reentering the Hubble horizon.
We consider the couplings that induce an effective mass of the background relativistic plasma, which change the overall equation of state of the Universe periodically and ultimately amplify curvature perturbations inside the Hubble horizon~\cite{Kapusta:1979fh,Quiros:1999jp}.
This amplification can be observed with the corresponding SGWB, which could be detected by multiband GW detectors. 
This mechanism is generally applicable, independent of a specific inflationary model or interaction term.


In Section~\ref{sec:light scalar}, we consider the evolution of the light scalar field and the impact on the sound speed of background radiation.
In Section~\ref{Parametric resonance}, we discuss the parametric resonance of curvature perturbations caused by the sound speed oscillations from the evolution of the light scalar field. We also establish the formulas for generating scalar-induced GWs.
In Section~\ref{sec:example}, we take the coupling of the light scalar field and electrons as an example, and find that the observation of such scalar-induced GWs can determine the coupling strength of the interaction and the initial value of the light scalar field.
We conclude in Section~\ref{sec:conclusion}.
For convenience, we set $c = \hbar = 1$ throughout this paper.

\section{Light Scalar Field with Quadratic Coupling} \label{sec:light scalar}

We consider the scenarios where the scalar field $\phi$ is charged under a $\mathcal{Z}_2$ symmetry~\cite{Damour:2010rm,Hees:2018fpg,Hees:2019nhn}, such that the primary coupling is quadratic, and the interaction term can be written as
\begin{equation}\label{eq:Vint}
    V_{\text{int}}= \begin{cases}
        \lambda^{2}\phi^2\psi^2, &\text{if}\, \psi \,\text{is a boson}\,,\\
        \lambda^{2}\phi^2\psi\bar{\psi}, &\text{if}\, \psi\, \text{is a fermion}\,.
    \end{cases}
\end{equation}
In this paper, we investigate the evolution of $\phi$ in the radiation-dominated~(RD) era and the backreaction of $\phi$ on the background plasma, and $\psi$ is a relativistic field that existed in the RD era, which can be the fields in the SM.
The parameter $\lambda$ determines the coupling strength between $\phi$ and $\psi$. We assume that the form of the interaction term~\eqref{eq:Vint} is valid during the RD era and $\lambda$ is a constant.
Moreover, we define $\mathcal{G}$ as the coupling constant between $\psi$ and thermal bath which is typically an SM gauge coupling, and also $\sigma\equiv\frac{\mathcal{G}^2}{4\pi}$.
In this paper, we assume that the interaction between $\psi$ and the thermal bath is much larger than the interaction between $\psi$ and the light scalar field and field value of $\phi$ is small, $\lambda\phi\ll T$.
Note that the light scalar field may interact with multiple fields at the same time and $V_{\mathrm{int}}$ represents the combined contribution of all fields that are quadratically coupled to $\phi$. Here, we consider the case that the coupling with $\psi$ is dominant. 

Under the high-temperature approximation, the finite temperature field theory yields the effective expression of the interaction term
\begin{equation}\label{eq:Veff}
    V_{\text{int}}=\frac{1}{2}f^{2}T^{2}\phi^{2}\,,
\end{equation}
where $f$ is a function of $\lambda$. In the high temperature approximation, $f$ is proportional to $\lambda$, $f=C\lambda$, where $C$ is a $\mathcal{O}(10)$ constant~\cite{Kapusta:2006pm,Kapusta:2023eix,Kapusta:1979fh,Quiros:1999jp}. 


The equation of motion~(EoM) of the light scalar field is given by
\begin{equation}\label{eq:EoM}
    \phi''+\left(2\mathcal{H}+\frac{2}{3}a\Gamma^{\text{eff}}_{\phi}\right)\phi'+m_{\text{eff}}^2a^2\phi=0\,,
\end{equation}
where a $prime$ denotes the derivative with respect to the conformal time $\eta$, $\mathcal{H}$ is the conformal Hubble parameter and $\Gamma^{\text{eff}}_{\phi}$ is the effective dissipative coefficient due to the dissipation effect through scattering processes. 
The effective mass, $m_{\mathrm{eff}}$, is a combination of the bare mass and the thermal mass induced by the coupling~\eqref{eq:Vint}, i.e., $m_{\text{eff}}^2=m_{\text{bare}}^2+m_{\text{ind}}^2$, where $m_{\text{ind}}^{2}\equiv \frac{d^{2}V_\text{int}}{d\phi^{2}}$. We are mainly concerned with the effects of the effective mass induced by the interaction, so in the following, we neglect the small bare mass of $\phi$.

Under the condition $\mathcal{H}/a\gg m_{\text{eff}}$, the damping term, $2\mathcal{H}\phi'$, dominates the EoM so that $\phi$ is almost a  constant.
Along with the expansion of the Universe, $\phi$ starts to oscillate around the minimum of its effective potential when $m_{\text{eff}}$ exceeds $\mathcal{H}/a$, which means that the light scalar field begins to oscillate with thermal mass.
Since we have assumed that $\lambda\ll \sigma$, the oscillation is adiabatic with respect to thermal relaxation rate and the dissipation is caused by scattering with $\psi$ particles in thermal bath, and the dissipation rate is evaluated as~\cite{Mukaida:2012qn}
\begin{equation}
    \Gamma^{\text{eff}}_{\phi}\sim\lambda^2\sigma T\,.
\end{equation}
Therefore, in the case of this work, $H$ is much larger than $\Gamma^{\text{eff}}_{\phi}$ so that the expansion of the Universe dominates the evolution of background scalar field $\phi$,
and the solution of Eq.~\eqref{eq:EoM} is obtained as




\begin{equation}
    \phi(\eta)=\phi_i\frac{\eta_i}{\eta}\cos{ (m_{\text{eff}}a(\eta-\eta_i))}\,,
\end{equation}
where $\eta_i$ satisfies $\mathcal{H}(\eta_i)=\frac{1}{\eta_i}\approx m_{\text{eff}}a(\eta_{i})$, representing the start time of the oscillation of $\phi$.

The oscillatory solution of $\phi$ will result in an oscillating effective mass of $\psi$ and the sound speed of background radiation.
The bare mass of $\psi$ is negligible compared to the temperature $T$ in the RD era.
Therefore, the effective mass of $\psi$ is approximately $M_{\text{eff}}\approx \lambda \phi(\eta)$, which is also induced from the interaction term~\eqref{eq:Vint}.
In general, it is very difficult to detect such weak interactions through experiments on the ground. We find the oscillations of the sound speed can largely amplify the scalar perturbations through parametric resonance, opening a new window to detect the dark sectors of our Universe.
The oscillation period of $\phi$ is comparable to the Hubble time, which is much larger than the relaxation time of the plasma. Therefore, the background radiation is in thermal equilibrium
\begin{equation}
\begin{split}
    \rho_{\psi}&=\frac{g_{\psi}}{2\pi^2}\int_{M_{\text{eff}}}^\infty\frac{\sqrt{\epsilon^2-M_{\text{eff}}^2}}{\exp(\frac{\epsilon}{T})\pm 1}\epsilon^2d\epsilon\\
    &=\frac{g_{\psi} T^4}{2\pi^2}\int_{\alpha}^\infty\frac{\sqrt{l^2-\alpha^2}}{\exp(l)\pm 1}l^2dl\,,
\end{split}
\end{equation}
\begin{equation}
    \centering
    \begin{split}
        P_{\psi}&=\frac{\rho_{\psi}}{3}-\frac{M^2_{\text{eff}}g_{\psi}}{6\pi^2}\int_{M_{\text{eff}}}^\infty\frac{\sqrt{\epsilon^2-M_{\text{eff}}^2}}{\exp(\frac{\epsilon}{T})\pm 1}d\epsilon\\
        &=\frac{\rho_{\psi}}{3}-\frac{\alpha^2g_{\psi}T^4}{6\pi^2}\int_{\alpha}^\infty\frac{\sqrt{l^2-\alpha^2}}{\exp(l)\pm 1}dl\,,
    \end{split}
\end{equation}
where $\alpha=\frac{M_{\text{eff}}}{T}$, $l=\frac{\epsilon}{T}$, $g_{\psi}$ is the effective degree of freedom of particle $\psi$, the signs $+$ and $-$ denote the fermion and boson case, respectively.
Because of the weakness of the interaction, we can safely adopt the approximation of $\alpha \ll 1$ and neglect the higher-order contributions of $\alpha$. Substituting the evolution of $\phi$, we can obtain the total energy density and pressure of radiation
\begin{equation}
    \rho_{\text{r}}=\frac{g\pi^4}{15}T^4-\frac{g_{\psi}}{12}\left(\frac{\lambda\phi_i\eta_i}{T\eta}\right)^2\frac{(T_i\eta_i)^4}{\eta^4}\cos^2\left(m_{\text{eff}}a(\eta-\eta_i)\right)\,,
\end{equation}
\begin{equation}
    P_{\text{r}}=\frac{g\pi^4}{45}T^4-\frac{g_{\psi}}{12}\left(\frac{\lambda\phi_i\eta_i}{T\eta}\right)^2\frac{(T_i\eta_i)^4}{\eta^4}\cos^2\left(m_{\text{eff}}a(\eta-\eta_i)\right)\,,
\end{equation} 
where $g$ is the effective degree of freedom of the relativistic fields in the early Universe.
Note that the temperature $T$ also oscillates with time. The evolution of $T$ satisfies the energy conservation. Since here we only focus on the sound speed of radiation which only depends on $M_{\mathrm{eff}}$, the evolution of $T$ has been omitted.
Under adiabatic approximation, the sound speed of radiation oscillates with a small amplitude,
\begin{equation}\label{eq:cs}
    c_s^2(\eta)=\frac{\partial P_{\text{r}}}{\partial\rho_{\text{r}}}\approx\frac{1}{3}-\frac{5}{12\pi^2}\frac{g_{\psi}}{g}\left(\frac{\lambda\phi_i}{T_i}\right)^2\cos(2k_0\eta-2k_0\eta_i)\,,
\end{equation}
where $k_0=m_{\text{eff}}(\eta)a(\eta)$ is a constant in the RD era. In what follows, we define $\Delta_s=\frac{g_{\psi}}{g}\left(\frac{\lambda\phi_i}{T_i}\right)^2$ for convenience.

\section{Parametric resonance and scalar-induced GWs}
\label{Parametric resonance}
The perturbed metric in the conformal Newtonian gauge reads
\begin{equation}
  ds^2=a(\eta)^2\left\{-(1+2\Phi)d\eta^2+\left[(1-2\Phi)\delta_{ij}+\frac{1}{2}h_{ij}\right]dx^idx^j\right\}\,,
\end{equation}
where we have neglected vector perturbations and the anisotropic stress.

The Universe is dominated by radiation, and the equation of motion for scalar perturbations can be written as
\begin{equation}
    \Delta\Phi-3\mathcal{H}\left(\Phi'+\mathcal{H}\Phi\right)=4\pi Ga^2\delta\rho_{\text{r}}\,.
\end{equation}
We apply the gauge-invariant variable $u=\exp(\frac{3}{2}\int(1+c_s^2)\mathcal{H}d\eta)\Phi$ so that the EoM of scalar perturbations in the Fourier space during the RD era reads~\cite{Mukhanov:2005sc}
\begin{equation}
    \label{evolution for scalar}
    u_k''+\left(c_s^2k^2-\frac{2}{\eta^2}\right)u_k=0\,.
\end{equation}

Together with Eq.~\eqref{eq:cs}, Eq.~\eqref{evolution for scalar} can be written in the form of the Mathieu equation
\begin{equation}
  \label{ME}
  \frac{du_k^2}{dx^2}+(A_k-2q \cos(2x))u_k=0\,,
\end{equation}
where $x=k_0\eta-k_0\eta_{i}$, $A_k=\frac{1}{3}{(\frac{k}{k_0})}^2-\frac{2}{x^2}$, $q=\frac{5}{24\pi^2}\Delta_s\left(\frac{k}{k_0}\right)^2$.
The parameter $A_k$ approaches the constant $\frac{1}{3}{(\frac{k}{k_0})}^2$ in the limit of large $x$.

Since $q$ is a small parameter, parametric resonance exists in the narrow ranges satisfying $A_k\sim N$, where $N$ is a positive integer.
In what follows, we focus on the dominant resonance band $1-q<A_k<1+q$. 
Because of the time dependence of $A_{k}$, the resonance band changes with time, and tends to stabilize to $\sqrt{\frac{3}{1+\frac{5}{8\pi^2}\Delta_s}}<\frac{k}{k_0}<\sqrt{\frac{3}{1-\frac{5}{8\pi^2}\Delta_s}}$.
The amplitude of the modes in the resonance band exponentially increases with time, $u_{k}\sim \exp(s_kx)$, with the factor
\begin{equation}
    \label{sk}
    s_k=\frac{q}{\sqrt{A_k}}\sqrt{1-\left(\frac{2\sqrt{A_k}(1-\sqrt{A_k})}{q}\right)^2}\,,
\end{equation}
which is mainly determined by $\phi_i$. 
At the beginning of the oscillation of $\phi$, $m_{\mathrm{eff}}$ is close to the Hubble parameter $\frac{\mathcal{H}}{a}$, so the energy density of $\phi$ is approximated as $\rho_{\phi}\approx\frac{1}{2}\left(\frac{\mathcal{H}}{a}\right)^{2}\phi_{i}^{2}$, which should be subdominant in the Universe.
The energy density fraction of $\phi$ is obtained as
\begin{equation}
    \Omega_{\phi}\equiv\frac{\rho_{\phi}}{\rho_{\text{tot}}}=\frac{4\pi Ga^2(\eta_{i})}{3}\frac{m_{\text{eff}}^2\phi_{i}^2}{\mathcal{H}^2(\eta_{i})}\,.
\end{equation}
which also yields
\begin{equation}
    \phi_{i}=\sqrt{\frac{3\Omega_{\phi}}{4\pi}}M_{\text{pl}}\,.
\end{equation}
indicating that $\phi_i$ much less than the Planck mass, $M_{\text{pl}}=1.221\times 10^{19}$ GeV.

Perturbations of energy density are amplified by parametric resonance and soon the linear perturbation approximation is violated. This provides the termination condition of parametric resonance. 
In the RD era, scalar metric perturbations, $\Phi$, are related to energy density perturbations, $\delta\rho$, as~\cite{Mukhanov:2005sc}
\begin{equation}
  \label{evolution of rho}
  \Delta\Phi-3\mathcal{H}(\Phi'+\mathcal{H}\Phi)=4\pi Ga^2\delta\rho\,.
\end{equation}
Perturbations of energy density should be smaller than the background energy density, which requires
\begin{equation}
    \label{cut off}
    \delta\rho(\mathbf{x},\eta)=\int\frac{d^3k}{{(2\pi)}^{\frac{3}{2}}}\delta\rho_k(\eta)\exp(i\mathbf{k}\mathbf{x})<\rho_0(\mathbf{x},\eta)\,,
\end{equation}
where $\rho_0$ represents the average background energy density, which is independent of the coordinate $\mathbf{x}$.
Eq.~\eqref{cut off} indicates that at any position $\mathbf{x}$ in the universe,  perturbations of the energy density cannot exceed the background value.

The amplified scalar perturbations can efficiently generate scalar-induced GWs which are expected to be observed by multiband GW observers. 
Scalar-induced GWs have received significant attention in recent years due to their ability as probes for detecting perturbations in the curvature of the early Universe~\cite{Inomata:2020yqv,Inomata:2018epa,Inomata:2019ivs,Inomata:2019zqy,Yuan:2019udt,Cai:2019elf,Yuan:2019wwo,Inomata:2019yww,Domenech:2019quo,Fu:2019vqc,Cai:2019bmk}.
In the following, we extract the primordial value $\phi_k$ from the definition of $\Phi_k(\eta)=T_k(k\eta)\phi_k$, ensuring that the transfer function $T_k(k\eta)$ approaches unity well before entering the Hubble horizon.
The primordial value $\phi_k$ is related to the power spectrum of curvature perturbations as
\begin{equation}
    \langle\phi_k\phi_{k'}\rangle=\delta(k+k')\frac{2\pi^2}{k^3}\left(\frac{3+3\omega}{5+3\omega}\right)^2\mathcal{P}_{\zeta}(k)\,.
\end{equation}

Most of the previous works focus on typical inflationary models that can generate large-amplitude curvature perturbations at small scales~\cite{Cai:2019bmk,Cai:2020ovp,Peng:2021zon}. Most Scalar-induced GWs are produced at the horizon-crossing time of amplified curvature perturbations. In contrast, we do not rely on the initial conditions of curvature perturbations determined by inflation, but simply apply $P_{\zeta}\sim 2\times 10^{-9}$ at small scales, which is the most natural prediction of inflation. The amplification of scalar perturbations is realized after reentering the horizon. The GWs production completes roughly around the time when $\delta\sim\delta\rho/\rho_{0}$ reaches the $\mathcal{O}(1)$ level. This is the main innovation point of this work.

As predicted by most well-motivated inflationary models, the power spectrum of curvature perturbations is approximately scale-invariant,
\begin{equation}
    \mathcal{P}_{\zeta}=A_{\zeta}\left(\frac{k}{k_*}\right)^{m_{\text{s}}-1}\,.
\end{equation}
From Planck-2018 data, we apply the scalar spectrum amplitude $A_{\zeta}=2.1\times10^{-9}$, pivot scale $k_*=0.05$ $\mathrm{Mpc}^{-1}$, and scalar spectral index  $m_{\text{s}}=0.965$~\cite{Planck:2018jri}.

The GW energy spectrum, $\Omega_{\text{GW}}(\eta,k)$, of the scalar-induced SGWB can be expressed as~\cite{Bouley:2022eer}
\begin{equation}
    \begin{split}
        \Omega_{\text{GW}}(\eta,k)=\frac{1}{6}n^2\int_0^{\infty}dv\int_{|1-v|}^{1+v}du\left(\frac{4v^2-(1+v^2-u^2)^2}{4uv}\right)^2\overline{I^2(v,u,n)}\mathcal{P}_{\zeta}(kv)\mathcal{P}_{\zeta}(ku)\,,
    \end{split}
\end{equation}
where we have introduced the dimensionless variable $n\equiv k\eta$, and the overline represents the average of the oscillations. 
The function $I(v, u, n)$ is defined as
\begin{equation}
\label{I}
    I(v,u,n)=\int_0^nd\overline{n}\frac{a(\overline{\eta})}{a(\eta)}kG_k(n,\overline{n})f(v,u,\overline{n})\,,
\end{equation}
where $G_k=\frac{1}{k}\sin(n-\overline{n})$ is the Green's function in the radiation-dominated Universe.

The source information is contained in the expression of $f(v,u,\overline{n})$
\begin{equation}
\label{fuv}
    \begin{split}
        f(v,u,\overline{n})&=\frac{6(\omega+1)}{3\omega+5}T(v\overline{n})T(u\overline{n})\\
        &+\frac{6(1+3\omega)(\omega+1)}{(3\omega+5)^2}\left(\overline{n}\partial_{\overline{\eta}}T(v\overline{n})T(u\overline{n})\right.\\
        &\left.+\overline{n}\partial_{\overline{\eta}}T(u\overline{n})T(v\overline{n})\right)\\
        &+\frac{3(1+3\omega)^2(1+\omega)}{(3\omega+5)^2}\overline{n}^2\partial_{\overline{\eta}}T(v\overline{n})\partial_{\overline{\eta}}T(u\overline{n})\,.
    \end{split}
\end{equation}

After some derivations~(see Appendix~\ref{gw equation} for the details), the energy spectrum of scalar-induced GWs in our scenario can be written in a simplified form
\begin{widetext}
    \begin{equation}
    \begin{split}
        \Omega_{\text{GW}}(n,k)=&0.0012A_{\zeta}^2T^2_{\text{end}}n^4_{\text{end}}\left(\frac{12k_0^2-k^2}{12k_0^2}\right)^2\left(\frac{k^2}{3k_0^2}\right)^2\\
        &\left(\mathcal{I}^2_1\left(\frac{\sqrt{3}k_0}{k},\frac{\sqrt{3}k_0}{k},n\rightarrow\infty,n_{\text{end}}\right)+\mathcal{I}^2_2\left(\frac{\sqrt{3}k_0}{k},\frac{\sqrt{3}k_0}{k},n\rightarrow\infty,n_{\text{end}}\right)\right)\left(\frac{\sqrt{3}k_0}{k_*}\right)^{2(m_{\text{s}}-1)}\Theta\left(2-\sqrt{3}\frac{k}{k_0}\right)\,.
    \end{split}
    \label{omega}
\end{equation}
\end{widetext}
where $n\equiv k\eta$, and $n_{\text{end}}$ and $T_{\text{end}}$ denote the value of $n$ and the temperature at the end of the resonance.
$\mathcal{I}_1$ and $\mathcal{I}_2$ are the equations in terms of $n_{\text{end}}$, $n$ and $k$, whose detailed form can be found in Eq.~\eqref{eq:I1I2} of Appendix \ref{gw equation}.
 As the intensity of the GW source quickly decreases with time, the GW energy density scales as radiation and the energy fraction of GWs stays almost constant during the RD era. 
 We can obtain the present energy spectrum of scalar-induced GWs through entropy conservation~\cite{Inomata:2020tkl,Espinosa:2018eve}
     \begin{equation}
        \Omega_{\text{GW},0}(k)h^2=0.39\left(\frac{g_*}{106.75}\right)^{-1/3}\Omega_{\gamma,0}h^2\Omega_{\text{GW}}(k)\,,
     \end{equation}
 where $\Omega_{\gamma,0}h^2\sim4.18\times10^{-5}$ is the energy density fraction of radiation at present~\cite{Planck:2018vyg}.


\section{results} \label{sec:example}

\begin{figure}
    \begin{minipage}[t]{0.49\linewidth}
        \includegraphics[width=3.5in]{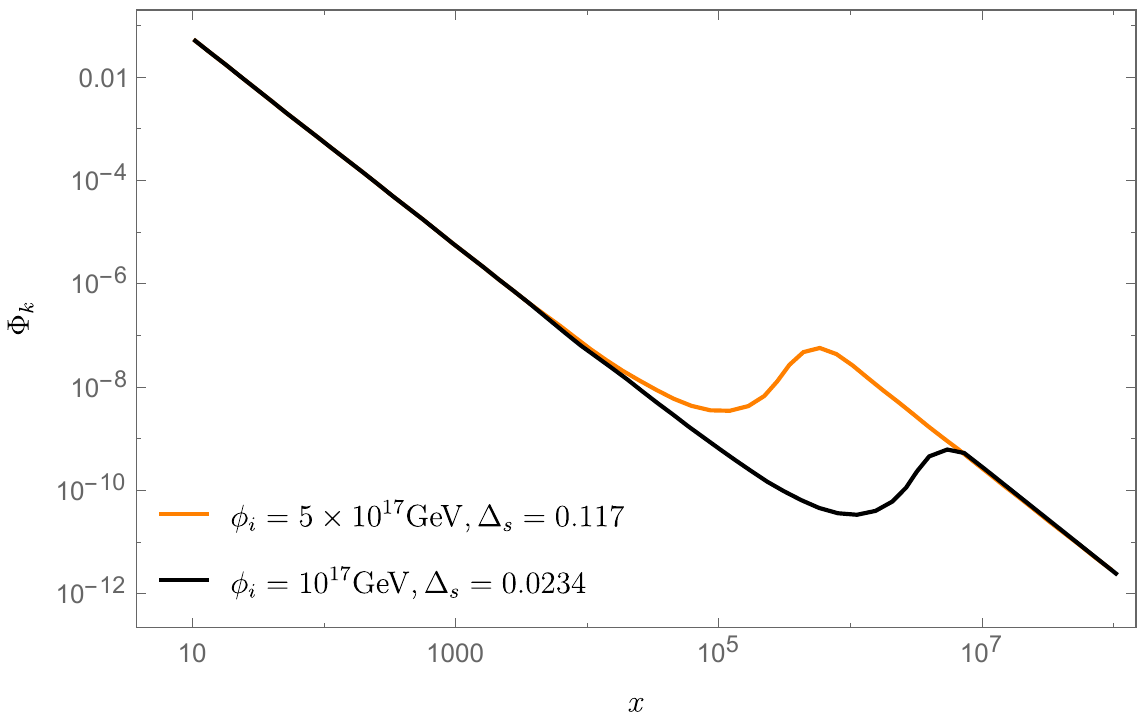}
    \end{minipage}
    \begin{minipage}[t]{0.49\linewidth}
        \includegraphics[width=3.5in]{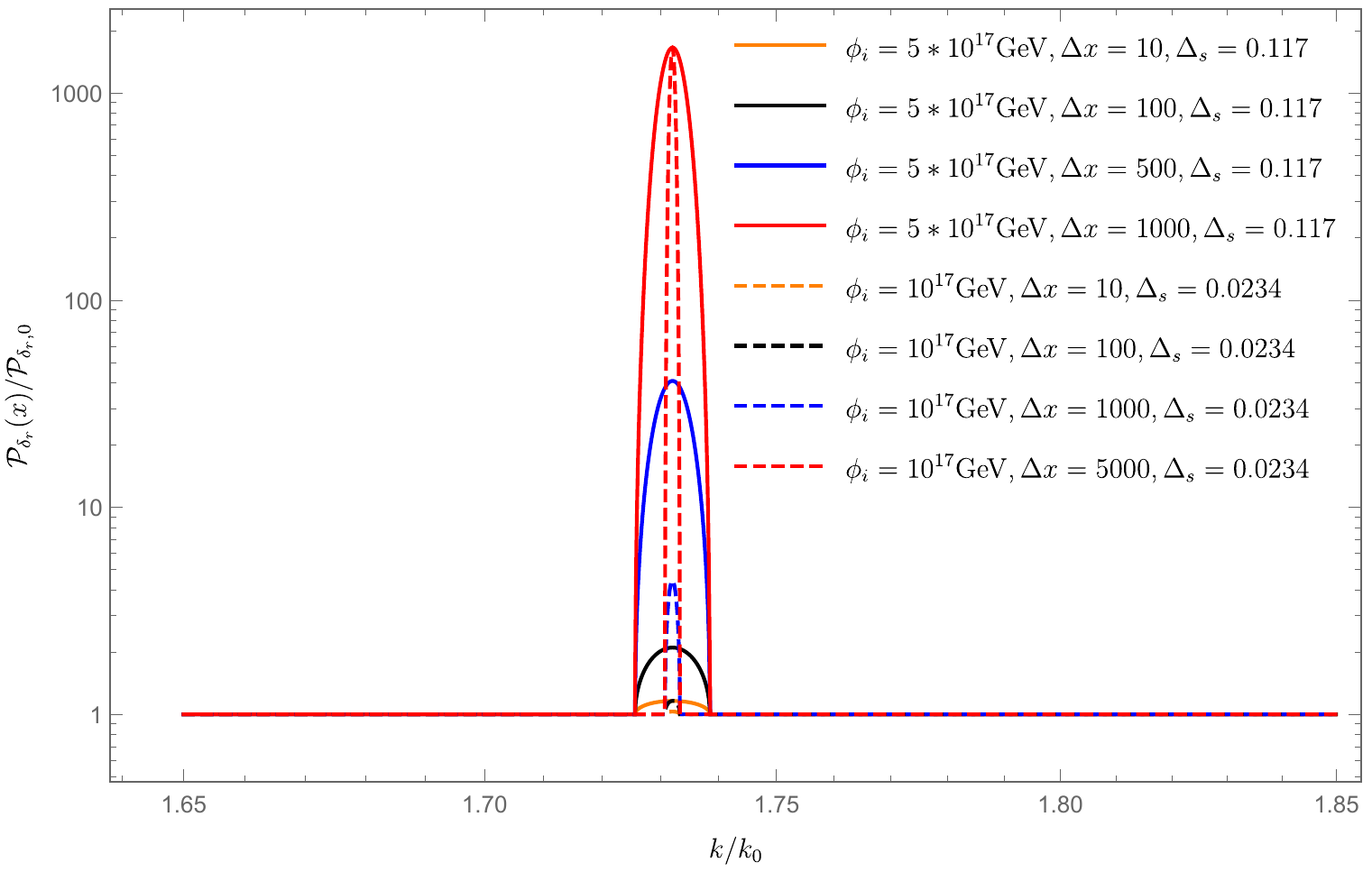}
    \end{minipage}
    \captionsetup{justification=raggedright}
    \caption{The left panel depicts the evolution of the amplitude of $\Phi_k$ in the instability band $k=\sqrt{3}k_0$ after reentering the Hubble horizon, where the orange line and the black line represent the cases with $\phi_i=5\times10^{17}$ $\mathrm{GeV}$ and $\phi_i=10^{17}$ $\mathrm{GeV}$, respectively. In the right panel, we show the power spectrum of density perturbations during the resonance with different $\phi_i$, where the solid line corresponds to $\phi_i=5\times10^{17}$ GeV, and the dashed line corresponds to $\phi_i=10^{17}$ GeV.}
    \label{phi}
\end{figure}

\begin{figure}
    \centering
    \includegraphics[width=5.0in]{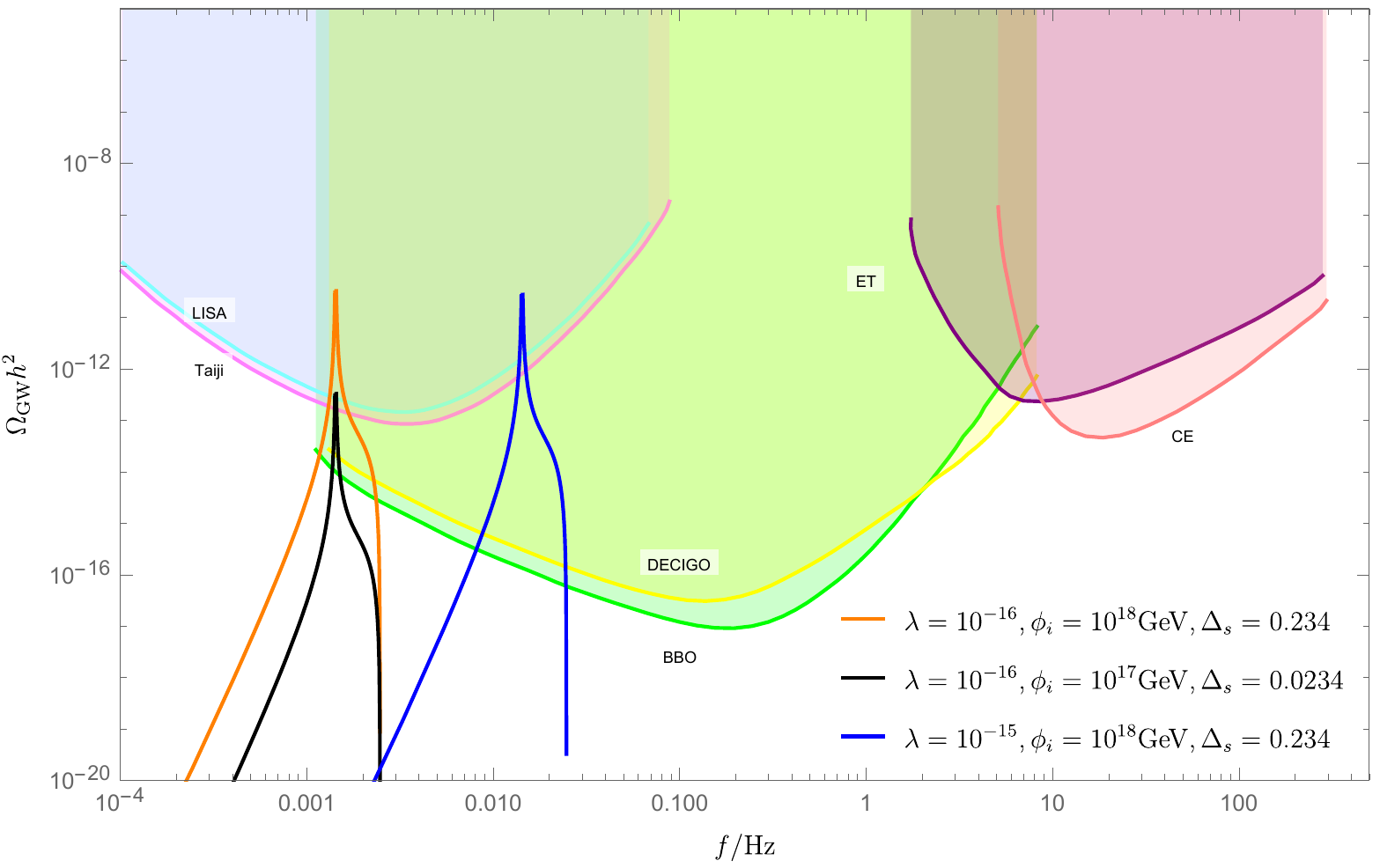}
    \captionsetup{justification=raggedright}
    \caption{The energy spectrum of scalar-induced GWs with $C=20$. As an illustration, We show the results with three parameter sets, $\lambda=10^{-16}$ with $\phi_i=10^{18}$ GeV, $\lambda=10^{-16}$ with $\phi_i=10^{17}$ GeV, and $\lambda=10^{-15}$ with $\phi_i=10^{18}$ GeV, shown in orange, black and blue lines, respectively.}
    \label{gwfig2}
\end{figure}

\begin{figure}
    \centering
    \includegraphics[width=5.0in]{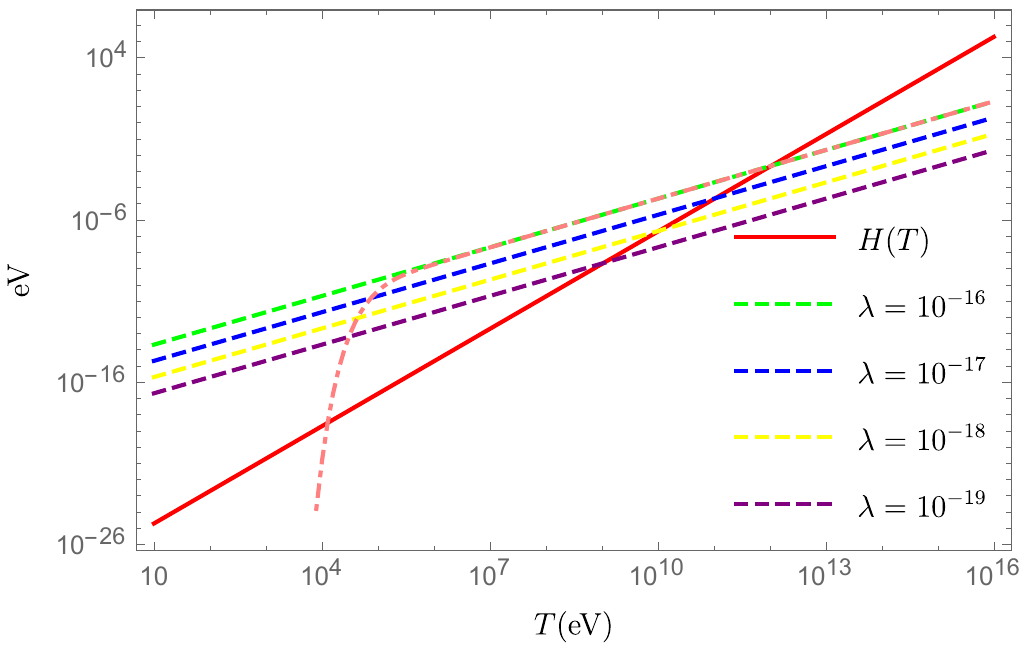}
    \captionsetup{justification=raggedright}
    \caption{The effective mass $m_{\mathrm{eff}}(\lambda)$ of $\phi$~(the dashed lines) with $\phi_i=10^{18}$ GeV, $\Delta_s=0.234$, and the Hubble parameter $H$~(the red solid line) in terms of the temperature of the Universe in the RD era.}  
    \label{H_and_meff}
\end{figure}

In this section, we $\frac{}{}$ present the numerical results of the evolution of $\Phi_k$ and the predictions of $\Omega_{\mathrm{GW,0}}(k)$.
The numerical results in Fig.~\ref{phi} and Fig.~\ref{gwfig2} show the evolution of scalar perturbations and the energy spectrum of scalar-induced GWs for different coupling constants $\lambda$ and initial values of the light scalar field $\phi_i$. The parameter $f$ in Eq.~\eqref{eq:Veff} is a function of $\lambda$, derived from the Matsubara sum and the propagator integration~\cite{Kapusta:2006pm,Kapusta:2023eix,Kapusta:1979fh,Quiros:1999jp}. 
Without loss of generality, we set $f=C\lambda$, where $C$ is a constant of the order $\mathcal{O}(10)$ depending on the specific coupling between $\phi$ and other matter fields.

In Fig.~\ref{phi}, we show the evolution of $\Phi_{k}$ with the resonant mode $k=\sqrt{3}k_0$. Here, $\Phi_{k}$ denotes its amplitude and we neglect the rapid oscillation behavior. 
In the resonance period, from $x_{\text{st}}$ to $x_{\text{end}}$, $\Phi_{k}$ are exponentially amplified by parametric resonance.  
Figure~\ref{phi} indicates that the start time of the resonance, $x_{\text{st}}$ becomes larger for a smaller $\phi_i$ while the total amplification rate of $\Phi_{k}$ is irrelevant to $\phi_{i}$. The evolution of $\Phi_{k}$ is roughly $\Phi_{k}(x)\sim x^{-2}e^{s_{k}x}$, the resonance begins to dominate the evolution at around the time $\frac{d\Phi_{k}}{dx}=0$, implying an estimation of $x_{\text{st}}\sim s_{k}^{-1}$. Equation~\eqref{sk} also implies $s_{k}$ is approximately proportional to $\phi_{i}$, which clarifies the reason why $x_{\text{st}}$ becomes larger for a smaller $\phi_{i}$. The total amplification rate of $\Phi_{k}$ is the same as $\delta\rho_{k}$, which is close to $10^{5}$, irrelevant to $\phi_{i}$.

Figure~\ref{gwfig2} shows the numerical results of scalar-induced GWs. Form this figure, we can see that the peak value of $\Omega_{\mathrm{GW}}$ is determined by $\phi_i$, while the peak frequency is determined by the $\lambda$. 
The reason is as mentioned before, a smaller $\phi_i$ corresponds to a smaller $s_{k}$ and a later resonance time $x_{\text{st}}$, reducing the time duration of GW generation. The resonant amplification occurs only for the modes with $k=\sqrt{3}k_0$. Since $k_{0}= fT(\eta)a(\eta)\approx C\lambda$, the peak frequency is roughly proportional to $\lambda$.

Note that all our discussions are conducted under the condition of high-temperature approximation, i.e., the fields that coupled to $\phi$ must be relativistic.

For a specific example, we consider the interaction between $\phi$ and electrons, which are motivated by both the dilaton and ultralight dark matter models~\cite{Damour:1994zq,Hui:2021tkt,Bouley:2022eer}.
\begin{equation}
    V_{\text{int}}= -\lambda^2\phi^2\overline{e}e\,.
\end{equation}
We utilize the finite-temperature field theory to obtain the effective mass contributed from the interaction term.
According to Matsubara's theory, the leading term of the self-energy reads
\begin{equation}
    \Pi_{\text{ele}}(\omega_m,p)=-2\lambda T\sum_n\int\frac{d^3\mathbf{p}}{(2\pi)^3}\frac{tr(\slashed p+m_e)}{p^2-m_e^2}\,,
\end{equation}
where $p_{\mu}$ is a four vector with $p_0=i(2n+1)T$, and we have defined that $p^2=p_{\mu}p^{\mu}$, $\slashed p=p_{\mu}\gamma^{\mu}$, and $\sum_n$ represents the summation over Matsubara frequencies. 
Up to the leading order, the temperature-dependent component of the mass reads 
\begin{equation}
\label{mind}
    m_{\text{ind}}^2=\lambda^2\frac{4}{\pi^2}T^2\int_{m_{\text{e}}/T}^\infty dx\,\frac{\sqrt{x^2-(m_{\text{e}}/T)^2}}{e^x+1}\,,
\end{equation}
which implies that $m_{\text{ind}}\propto T$ For $T\gg m_e$. Figure~\ref{H_and_meff} illustrates the dependence of $H$ and $m_{\mathrm{eff}}$ on the temperature $T$, where the homogeneous field $\phi$ begins to oscillate when the effective mass $m_{\mathrm{eff}}$ exceeds the Hubble parameter $H(T)$~\cite{Mukhanov:2005sc}.


The red dashed line in Fig.~\ref{H_and_meff} shows that the high-temperature approximation might be violated with the decrease of the temperature of the Universe. In this case, the resonance ceases before $\delta\sim \mathcal{O}(1)$ which reduces the 
amplification rate of scalar perturbations and the peak value of $\Omega_{\mathrm{GW}}$.

\section{Conclusions} \label{sec:conclusion}

In this paper, we present a novel approach to detect the weak quadratic coupling between a homogeneous light scalar field $\phi$ and background radiation through the observations of SGWBs. The coupling term induces effective masses for both $\phi$ and background radiation, resulting in the oscillation of $\phi$ and periodical variation of the speed of sound of the background plasma. 
The sound speed oscillations lead to the parametric amplification of scalar perturbations inside the Hubble horizon. The parametric resonance ceases when energy density perturbations are amplified to the $\mathcal{O}(1)$ level. We find the amplified scalar perturbations induce an observable SGWB which is expected to be detected in various GW observers. The peak frequency and the intensity of the SGWB respectively imply the coupling coefficient $\lambda$ and the initial value of the light scalar field, as shown in Fig.~\ref{gwfig2}. 



Although we only discuss electromagnetic interactions as an example, we acknowledge that the light scalar field can interact with radiation in multiple ways simultaneously.
So this mechanism can be used as a way to detect new physics.
In this case, we need to systematically analyze the contributions of various interactions to the effective mass, considering their individual cutoff conditions.
We directly take $g$ as a constant for simplicity. In a more realistic case, $g$ might gradually decrease with the expansion of the Universe, which will further enhance the strength of GWs.

Furthermore, we are disregarding the contribution of scalar static mass.
If the bare mass is not negligible, the resonance amplification band shifts over time.
In this case, we need to introduce new resonance cutoff conditions when the bare mass becomes nonnegligible.

\begin{acknowledgments}
    This work is supported in part by the National Key Research and Development Program of China Grants No. 2020YFC2201501 and No. 2021YFC2203002, in part by the National Natural Science Foundation of China Grants No. 12105060, No. 12147103, No. 12235019, No. 12075297 and No. 12147103, in part by the Science Research Grants from the China Manned Space Project with NO. CMS-CSST-2021-B01,	in part by the Fundamental Research Funds for the Central Universities. X.-Y.~Y. is supported in part by the KIAS Individual Grant No. QP090701.
\end{acknowledgments}

\appendix
\section{The calculation of scalar-induced GWs}
\label{gw equation}

\begin{figure}
    \centering
    \includegraphics[width=5.0in]{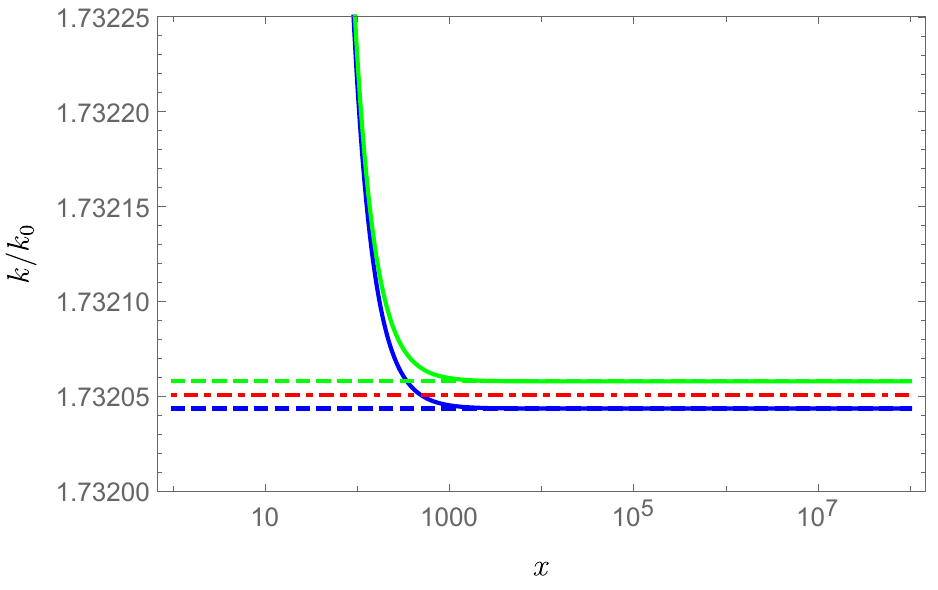}
    \caption{The evolution of the first resonance band with $\lambda=10^{-15}$ and $\phi_i=5\times10^{17}$ $\mathrm{GeV}$.
    The blue solid line and the green solid line represent the time for the $k$-mode perturbations to enter and leave the resonance band, respectively.}
    \label{bandfig}
\end{figure}

In the appendix, we derive the formulas used to calculate Eq.~\eqref{I}.
In Fig.~\ref{bandfig}, we show the evolution of the resonance band.
As the Universe expands, the resonance band gradually tends to stabilize. Because of the smallness of $\Delta_s$, the stable resonance band is very narrow so that we can approximately assume that the amplified power spectrum from parametric resonance is monochromatic with the mode $k=\sqrt{3}k_{0}$. 

For the modes out of the resonance band, Eq.~\eqref{evolution for scalar} gives the approximate solution of $T_k$ in the RD era
\begin{equation}
    T_k(n)=\frac{9}{n^2}\left(\frac{\sin\left(n/\sqrt{3}\right)}{n/\sqrt{3}}-\cos\left(n/\sqrt{3}\right)\right)\,,
\end{equation}
which implies $\Phi_k$ scales as $n^{-2}$ for large $n$. When the modes enter the resonance band at $n_{\mathrm{st}}$, the amplitude of curvature perturbations begins to increase exponentially as $T_k\sim\frac{\exp\left(s_k(n-n_{\mathrm{st}})\right)}{n^2}$.

The parametric resonance terminates when $\frac{\delta\rho}{\rho}\sim\mathcal{O}(1)$, and the end time $n_{\text{end}}$ is determined by the condition Eq.~\eqref{cut off}.

For $n>n_{\text{end}}$, the equation for the transfer equation with boundary conditions in the RD era reads
    \begin{equation}
    \begin{cases}
         &T_k''+\frac{4}{\eta}T_k'+\frac{1}{3}k^2T_k=0\,. \\
         &T_k(n_{\text{end}})=T_{k,\text{end}}\,,\\
         &T_k'(n)|_{n-n_{\text{end}}}=0\,.
    \end{cases}
    \end{equation}

Therefore, the transfer equation for $n>n_{\text{end}}$ can be written as 
\begin{widetext}
    \begin{equation}
    T_k(n)=\frac{\delta\left(k-\sqrt{3}k_0\right)}{n^2}\left(A\left(\frac{\sin \left(n/\sqrt{3}\right)}{n/\sqrt{3}}-\cos\left(n/\sqrt{3}\right)\right)+B\left(\frac{\cos \left(n/\sqrt{3}\right)}{n/\sqrt{3}}+\sin\left(n/\sqrt{3}\right)\right)\right)\,,
\end{equation}
\end{widetext}
where the factor $\frac{1}{\sqrt{3}}$ denotes the sound speed in the RD era, $A$ and $B$ are parameters that depend on the boundary conditions
\begin{equation}
A=T_{k,\text{end}}n_{\text{end}}^2\frac{\sin \frac{n_{\text{end}}}{\sqrt{3}}}{-\cos^2\frac{n_{\text{end}}}{\sqrt{3}}+\sin^2\frac{n_{\text{end}}}{\sqrt{3}}}=T_{k,\text{end}}n_{\text{end}}^2a\,,
\end{equation}
\begin{equation}
B=T_{k,\text{end}}n_{\text{end}}^2\frac{-\cos \frac{n_{\text{end}}}{\sqrt{3}}}{-\cos^2\frac{n_{\text{end}}}{\sqrt{3}}+\sin^2\frac{n_{\text{end}}}{\sqrt{3}}}=T_{k,\text{end}}n_{\text{end}}^2b\,.
\end{equation}
Here we have used the approximation of $n_{\text{end}}\gg 1$, we define two new variables, $a$ and $b$, which depend only on $n_{\text{end}}$ for convenience.
The evolution of scalar perturbations can be found in Fig.~\ref{phi}.
It should be noted that, as found by Eq.~\eqref{I} and Eq.~\eqref{fuv}, when computing the kernel term $I(u,v,n)$, it is mainly calculated for time, so we leave the $\delta(k-\sqrt{k_0})$ term out of consideration for the time being until we are ready to integrate $u$ and $v$.

After the end of the parametric resonance, i.e., $n>n_{\text{end}}$, the evolution of scalar perturbations returns to the scaling solution $T_{k}\sim n^{-2}$. 
The total energy spectrum of scalar-induced GWs is contributed from three stages: before, during, and after the resonance. We find that the GW production rate of the first two stages is negligible and only calculate the contribution from the last stage for simplicity 


\begin{equation}
    \begin{split}
         I(v,u,n,n_{\text{end}})&\approx\frac{\sin n}{n}\int_{n_{\text{end}}}^n d\overline{n}\,\overline{n}\cos\overline{n}f(v,u,\overline{n})\\
         &-\frac{\cos n}{n}\int_{n_{\text{end}}}^nd\overline{n}\,\overline{n}\sin\overline{n}f(v,u,\overline{n})\,.
    \end{split}
\end{equation}

Since $n_{\text{end}}\gg1$, the source function $f$ is approximately
\begin{widetext}
    \begin{equation}
        f\left(v,u,\overline{n},n_{\text{end}}\right) \approx\frac{0.12}{uv}\left(T_{end}n_{\text{end}}^2\right)^2\frac{1}{\overline{n}^2} 
        \left(a^2\sin\frac{u\overline{n}}{\sqrt{3}}\sin\frac{v\overline{n}}{\sqrt{3}}+b^2\cos\frac{u\overline{n}}{\sqrt{3}}\cos\frac{v\overline{n}}{\sqrt{3}}-ab\sin\frac{\left(u+v\right)\overline{n}}{\sqrt{3}}\right)\,.
    \end{equation}
\end{widetext}

To calculate $I(v, u, n, n_{\text{end}})$, it is necessary to utilize the trigonometric addition theorem and integration by parts multiple times~\cite{Ananda:2006af}
\begin{widetext}
    \begin{equation}\label{eq:I1I2}
    \begin{split}
         I(u,v,n,n_{\text{end}})&\approx\frac{0.12\left(T_{\text{end}}n^2_{\text{end}}\right)}{uv}\frac{1}{n}\Bigg(\sin n\bigg(\frac{a^2-b^2}{4}\Big(\mathrm{Ci}(\mathrm{D}_{++}n_{\text{end}})+\mathrm{Ci}(\mathrm{D}_{+-}n_{\text{end}})-\mathrm{Ci}(\mathrm{D}_{++}n)-\mathrm{Ci}(\mathrm{D}_{+-}n)\Big)\\
    &\frac{-a^2-b^2}{4}\Big(\mathrm{Ci}(\mathrm{D}_{-+}n_{\text{end}})+\mathrm{Ci}(\mathrm{D}_{--}n_{\text{end}})-\mathrm{Ci}(\mathrm{D}_{-+}n)-\mathrm{Ci}(\mathrm{D}_{--}n)\Big)\\
    &\frac{ab}{2}\Big(\mathrm{Si}(-\mathrm{D}_{++}n_{\text{end}})-\mathrm{Si}(\mathrm{D}_{+-}n_{\text{end}})+\mathrm{Si}(\mathrm{D}_{++}n)+\mathrm{Si}(\mathrm{D}_{+-}n)\Big)\bigg)\\
    &-\cos n\bigg(\frac{b^2-a^2}{4}\Big(-\mathrm{Si}(\mathrm{D}_{++}n_{\text{end}})+\mathrm{Si}(\mathrm{D}_{+-}n_{\text{end}})+\mathrm{Si}(\mathrm{D}_{++}n)-\mathrm{Si}(\mathrm{D}_{+-}n)\Big)\\
    &+\frac{a^2+b^2}{4}\Big(-\mathrm{Si}(\mathrm{D}_{-+}n_{\text{end}})-\mathrm{Si}(\mathrm{D}_{--}n_{\text{end}})+\mathrm{Si}(\mathrm{D}_{-+}n)+\mathrm{Si}(\mathrm{D}_{--}n)\Big)\\
    &+\frac{ab}{2}\Big(\mathrm{Ci}(\mathrm{D}_{++}n_{\text{end}})-\mathrm{Ci}(\mathrm{D}_{+-}n_{\text{end}})-\mathrm{Ci}(\mathrm{D}_{++}n)+\mathrm{Ci}(\mathrm{D}_{+-}n)\Big)\bigg)\Bigg)\\
    &=\frac{0.12\left(T_{\text{end}}n^2_{\text{end}}\right)}{uv}\frac{1}{n}\big(\sin n\,\mathcal{I}_1(u,v,n,n_{\text{end}})-\cos n\,\mathcal{I}_2(u,v,n,n_{\text{end}})\big)\,,
    \end{split}
    \end{equation}
\end{widetext}
where the last line gives the definition of $\mathcal{I}_{1}$ and $\mathcal{I}_{2}$, $\mathrm{D}_{\pm}\equiv\frac{u\pm v\pm \sqrt{3}}{\sqrt{3}}$, and $\mathrm{Si}(n)$ and $\mathrm{Ci}(n)$ functions are defined as 
\begin{equation}
    \mathrm{Si}(n)=\int_0^nd\overline{n}\,\frac{\sin\overline{n}}{\overline{n}}\,,
\end{equation}
\begin{equation}
    \mathrm{Ci}(n)=-\int_n^\infty d\overline{n}\,\frac{\cos\overline{n}}{\overline{n}}\,.
\end{equation}

For the case $\eta\rightarrow\infty$, and consider the $\delta$ term,we can obtain the oscillation average to simplify the result
\begin{equation}
    \overline{I^2(u,v,n,n_{\text{end}})}=0.0072\delta(ku-\sqrt{3}k_0)\delta(kv-\sqrt{3}k_0)\frac{T^2_{\text{end}}n^4_{\text{end}}}{u^2v^2n^2}\left(\mathcal{I}_1^2(u,v,n\rightarrow\infty,n_{\text{end}})+\mathcal{I}_2^2(u,v,n\rightarrow\infty,n_{\text{end}})\right)\,.
\end{equation}

Then, the GW energy spectrum reads

\begin{widetext}
    \begin{equation}
    \begin{split}
        \Omega_{\text{GW}}(n,k)=&\frac{0.0072A_{\zeta}^2T^2_{\text{end}}n^4_{\text{end}}}{6}\int_0^{\infty}\int_{|1-v|}^{1+v}du\,\delta(ku-\sqrt{3}k_0)\delta(kv-\sqrt{3}k_0)\left(\frac{4v^2-\left(1+v^2-u^2\right)}{4uv}\right)^2\\
        &\frac{\mathcal{I}_1^2(u,v,n\rightarrow\infty,n_{\text{end}})+\mathcal{I}_2^2(u,v,n\rightarrow\infty,n_{\text{end}})}{u^2v^2}\left(\frac{k^2uv}{k^2_*}\right)^{m_s-1}\,.
    \end{split}
\end{equation}
\end{widetext}
In this monochromatic case, the GW energy spectrum is
\begin{widetext}
    \begin{equation}
    \begin{split}
        \Omega_{\text{GW}}(n,k)=&0.0012A_{\zeta}^2T^2_{\text{end}}n^4_{\text{end}}\left(\frac{12k_0^2-k^2}{12k_0^2}\right)^2\left(\frac{k^2}{3k_0^2}\right)^2\\
        &\left(\mathcal{I}^2_1\left(\frac{\sqrt{3}k_0}{k},\frac{\sqrt{3}k_0}{k},n\rightarrow\infty,n_{\text{end}}\right)+\mathcal{I}^2_2\left(\frac{\sqrt{3}k_0}{k},\frac{\sqrt{3}k_0}{k},n\rightarrow\infty,n_{\text{end}}\right)\right)\left(\frac{\sqrt{3}k_0}{k_*}\right)^{2(m_{\text{s}}-1)}\Theta\left(2-\sqrt{3}\frac{k}{k_0}\right)\,.
    \end{split}
    \label{omega}
\end{equation}
\end{widetext}
\bibliography{citeLib}

\end{document}